\documentclass{ppig}
\usepackage{epsfig}
\usepackage{booktabs}
\usepackage{ucs}
\usepackage[utf8x]{inputenc}

\title{The Systems Approach to Change and the Agile Software Development Context}

\author{Lucas Gren \\
  Chalmers and the University of Gothenburg\\
  Gothenburg, Sweden 412--92\\
  lucas.gren@cse.gu.se \\
}

\date{}

\begin{document}
\maketitle
\thispagestyle{empty}

\begin{abstract}
There is a diversity of models explaining organizational culture and how these complex aspects can be addressed in connection to organizational change efforts. This workshop paper claims that models already exist for dealing with the cultural change that an agile transition is in the software engineering context. Instead of realizing this again through agile success stories, and thus reinventing the wheel, it is argued that the research in the software engineering field should build on these models instead and investigate how\slash if they differ. Practitioners already work as the change agents described in other fields and they should get recognition through the presence and integration of these models in the software engineering process research. 
\end{abstract}

\section{Introduction}
A few studies set out to investigate the social or cultural aspects of agile development. \citeA{whit}, for example, verifies that agile teams need to look at social-psychological aspects to fully understand how they function. There are also a set of studies connecting agile methods to organizational culture. These connect the agile adoption process to culture to see if there are cultural factors that could jeopardize the agile implementation, which there are \cite{iivari,tolfo2008}. One study divides culture in different layers depending on their visibility according to \citeA{scheincultlead}. This article shows that an understanding of culture layers increase the understanding of how an agile culture could be established \cite{tolfo}. However, the research within this sub-field lacks an analysis what happens if a hard (process) solution is used for soft (cultural) problems, something that needs to be realized in research on agile software development and through that give credibility to practitioners working on these aspects. Strategies already exist in other fields that could be directly implemented in the software engineering context and research can then focus on specific areas where the software engineering field is different and not reinvent the wheel.


\subsection{Organizational Culture and the Iceberg}\label{sec:iceberg}
A well established model is to view organizational culture as an iceberg. Above the surface is the external (surface) culture and is only representing 10\% of the organizational culture. These behaviors, traditions, customs, and structures are explicitly learned, conscious, easily changeable and constitute objective knowledge. However, 90\% of the culture (which is often ignored to a large extent) is below the surface. Here we have habits, core values, beliefs, priorities, politics, attitudes, perceptions, and assumptions which are all implicitly learned, unconscious, difficult to change, and constitute subjective knowledge \cite{french}. Most of the organizational change initiatives fail due to lack of these under-the-surface aspects of an organization of people \cite{strebel}.

\subsection{Agile Methods Over and Under the Surface}\label{sec:discussion}
The organizational (or cultural) iceberg metaphor is highly relevant for agile management just like any other human endeavor. To only focus on process, no matter if it is on waterfall methods or agile practices, is to only look at the peak of the iceberg. There are some research done in agile software development that point to this mistake (see e.g.\ \citeA{doingtobeing}), but also a study showing the cultural anchoring needed when implementing such practices successfully \cite{sharp}. The agile community often separate agile principles from agile practices. The principles are the values and culture of being agile and is often directly cited. These have become a popular way of defining ``agility'' and are as follows (http://agilemanifesto.org/): (1) Our highest priority is to satisfy the customer through early and continuous delivery of valuable software. (2) Welcome changing requirements, even late in development. Agile processes harness change for the customer's competitive advantage. (3) Deliver working software frequently, from a couple of weeks to a couple of months, with a preference to the shorter timescale. (4) Business people and developers must work together daily throughout the project. (5) Build projects around motivated individuals. Give them the environment and support they need, and trust them to get the job done. (6) The most efficient and effective method of conveying information to and within a development team is face-to-face conversation. (7) Working software is the primary measure of progress. (8) Agile processes promote sustainable development. The sponsors, developers, and users should be able to maintain a constant pace indefinitely. (9) Continuous attention to technical excellence and good design enhances agility. (10) Simplicity --the art of maximizing the amount of work not done-- is essential. (11) The best architectures, requirements, and designs emerge from self-organizing teams. (12) At regular intervals, the team reflects on how to become more effective, then tunes and adjusts its behavior accordingly.

With the lens of the iceberg metaphor it is clear that many of these principles are a dive under the surface. This metaphor makes it clear what the differences are between agile principles (being agile) and agile practices (doing agile). This is also connected to the lack of need for agile maturity models for practitioners \cite{xp20141}. \citeA{tolfo} also found it useful to sort organizational observations to further understand agile transformations in companies. Practitioners need tools for dealing with culture and not only structures for measuring it. The more formal methods focus on top-of-the-iceberg aspects whilst other models (like \citeA{sidky}) blend agile principles and practices in their assessment. Also, according to \citeA{williams}, 64.6\% of 326 experienced agile practitioners stated that the reason why agile principles are valuable is: ``because all agile teams choose among software development practices, but, if they want to be agile, they should choose practices that are in line with the principles''. So we need to start with a below-the-surface plan and then select practices to support those organizational changes. If we try to change the culture by changing visible processes (or in other words, using ``hard'' solutions for ``soft and messy'' problems) we will surely fail \cite{strebel}. We will now describe what this means in more detail.

\section{Discussion on Hard and Soft Solutions in Connection to Agile Development}\label{sec:hardsoft}
There are mainly two different aspects of any organizational change, the content of change and the context where it happens. The management ideas are often believed to be generic and we are taught to see the similarities of all types of organizations, instead of their differences. We think of a world full of organization instead of unique operative units. In order to translate organizational ideas they have to be ``decontextualized'' and contextualized again in a new organization. A key to implement new management ideas is to have what \citeA{managebook} calls translator skills. One must have knowledge about the context in which one tries to implement new methods. A great problem when using generic methods is that focus always is on mean value of success. This knowledge says very little about how well a method works in one specific case. This is also a reason why organizations often measure their agility in their own adapted way, i.e.\ the measurement models do not take the context into account enough \cite{jalali2014}.

What is a hard solution to a soft problem? In order to understand the effect of this, we must define what the differences are. The spectrum and difference are well described by \citeA{paton2008} and can be seen in Table~\ref{tropics}.
\begin{table}[h]
\caption{The TROPICS test.}
	\begin{center}
		\begin{tabular}{c||c||c}
\hline
\bfseries TROPICS factor & \bfseries ``Hard'' problem & \bfseries ``Soft'' problem \\
\hline\hline
Timescales & Clearly defined: short to medium term & Ill defined: medium to long term \\
\hline
Resources & Clearly defined and reasonably fixed & Unclear and variable \\
\hline
Objectives & Objective and quantifiable & Subjective and visionary \\
\hline
Perceptions & Shared by those affected  & Creates conflict of interest \\
\hline
Interest & Limited and well defined & Widespread and ill defined \\
\hline
Control & Within the managing group & Shared out with the group \\
\hline
Source & Originates internally & Originates externally \\
\hline
\end{tabular}

\label{tropics}
	\end{center}
\end{table}
In fact, a hard problem solution is based on systems engineering and older management ideas \cite{paton2008} just like traditional waterfall methods. A software engineering process change is far from that clear to the organization. Also, the cultural changes described in the Agile Manifesto and research conducted by e.g.\ \citeA{doingtobeing,tolfo} shows that agility is undeniably a soft issue as well as hard. 

Hard problems solution can be, for example, a simple change in an IT support system used by employees or acquisition of equipment or maintenance of supplies needed for the work place. A hard systems methodology of change usually has a description phase (with an analysis of the situation, identification of objective and constraints, and how to measure a successful change are included), options phase (evaluate different option compared to the performance measure), and finally, an implementation phase (carry out and evaluate the changes with given measures). 

The solution to soft problems can be called an OD (Organizational Development) approach. An almost too accurate description of OD to agile development processes was written by \citeA{french} (first edition out in 1973!). They describe the OD process as: ``A long term effort, led and supported by top management, to improve an organization's visioning, empowerment, learning and problem-solving processes, through an ongoing, collaborative management of organization culture -- with special emphasis on the culture of intact work teams and other team configurations -- using the consultant-facilitator role and the theory of technology of applied behavioral science.'' All the agile development lessons learned from success stories the last decade are there. The long term cultural effort, the executive buy-in, the empowered team members, the collaborative team environment, the facilitator role, and the recent findings of the usefulness of applied behavioral science \cite{iivari,tolfo2008}. 

The OD approach is considering the whole system as well as its parts. 
The first step is then to diagnose the current situation with regards to organizational purpose, goals, structure, culture, prevailing leadership approaches and styles, recruitment practices, career paths and opportunities, reward structures and practices, individuals' motivation and commitment to their work and organization, employee training and development provision, intra- and inter-group relationships etc. The second step is to develop a vision for change. One does not convince people without meeting them or showing them why our change is important. The vision is the core values and the end goal of the change (the desired state). After this, there is a substantial work with gaining commitment to the vision and the need for change. This is where the hard solution approach fail with devastating consequences. Unless concerned -- and to be involved in the process -- are consulted and have been a part of the creation of the vision they will have little incentives for ``buy-in''. At this state one can not communicate too much because the more information the group members get of what is going on the more will the back up the process. One common mistake is to focus on the people that are against you, or the ones that are not yet convinced. The key to a cultural change is to instead focus on the people that are with you and let them be bearers of the change culture. The forth step is to develop an action plan and have change agents that help the process. In the successful agile examples these change agents (that help with the change and let managers focus on day-to-day issues more) are often called ``agile coaches''. To have a process change facilitator is key to success in OD. The fifth step is to implement the change, assess it, and reinforce it. The later means that the change needs to be institutionalized to be long-term. 


If there is too much focus on process (no matter if it is on the new agile process or the old waterfall method) the initial problem remains. Replacing a waterfall model by a shallow agile set of practices will not work and the problem has then not been that one did not follow the agile process as strictly as one probably needed. We believe the idea about building flexibility into a process is an old idea from other fields and that agility is about implementing a responsive culture and not about the practices, which are only enablers for responsiveness and not the responsiveness itself. 


A reason for why a focus on above the water surface of the iceberg could be that it would seem safer for engineers to work on these concrete areas of concern, which makes the work for the OD change agent more difficult in the IT field. They resemble more the simplified and deterministic systems they were taught in their education \cite{seovercoming}. One has to be braver when investigating motives, values, cultures, etc.\ of people in an organization, especially with little training in these topics. This is also why intelligent technical experts sooner or later write books on behavioral subjects (like, for example, coaching \cite{adkins}) that are well-known facts to people with a behavioral science background. 

The software engineering field should realize that the idea of agile development processes' responsiveness and flexibility and their complex implementation are not new aspects of human or organizational behavior and theorize around these concepts in the light of what is already know in other fields. That way, the field can focus on researching how\slash if they differ instead of making the same mistakes others did decades ago.



\bibliography{references}
\bibliographystyle{apacite} 
\end{document}